%% file: arxiv.tex
\documentclass[sigconf]{acmart}
\acmSubmissionID{155}

\usepackage{booktabs} 
\usepackage{tabularx}
\usepackage{multirow}
\usepackage{siunitx}
\usepackage{array}
\usepackage{gensymb}

\newcolumntype{L}[1]{>{\raggedright\let\newline\\\arraybackslash\hspace{0pt}}m{#1}}
\newcolumntype{C}[1]{>{\centering\let\newline\\\arraybackslash\hspace{0pt}}m{#1}}
\newcolumntype{R}[1]{>{\raggedleft\let\newline\\\arraybackslash\hspace{0pt}}m{#1}}

\usepackage[ruled]{algorithm2e} 

\SetAlFnt{\small}
\SetAlCapFnt{\small}
\SetAlCapNameFnt{\small}
\SetAlCapHSkip{0pt}

\copyrightyear{2023}
\acmYear{2023}
\setcopyright{rightsretained}
\acmConference[SA Conference Papers '23]{SIGGRAPH Asia 2023 Conference Papers}{December 12--15, 2023}{Sydney, NSW, Australia}
\acmBooktitle{SIGGRAPH Asia 2023 Conference Papers (SA Conference Papers '23), December 12--15, 2023, Sydney, NSW, Australia}
\acmDOI{10.1145/3610548.3618136}
\acmISBN{979-8-4007-0315-7/23/12}

\begin{document}
\title{Drivable Avatar Clothing: Faithful Full-Body Telepresence with Dynamic Clothing Driven by Sparse RGB-D Input}

\author{Donglai Xiang}
\email{donglaix@cs.cmu.edu}
\affiliation{%
 \institution{Carnegie Mellon University}
 \country{USA}
}

\author{Fabian Prada}
\email{fabianprada@meta.com}

\affiliation{%
 \institution{Meta Reality Labs Research}
 \country{USA}
}

\author{Zhe Cao}
\email{zhecao@berkeley.edu}

\affiliation{%
 \institution{Meta Reality Labs Research}
 \country{USA}
}

\author{Kaiwen Guo}
\email{guokaiwen_neu@126.com}
\affiliation{%
 \institution{Meta Reality Labs Research}
 \country{USA}
}

\author{Chenglei Wu}
\email{chengleiwu@gmail.com}
\affiliation{%
 \institution{Meta Reality Labs Research}
 \country{USA}
}

\author{Jessica Hodgins}
\email{jkh@cmu.edu}
\affiliation{%
 \institution{Carnegie Mellon University}
 \country{USA}
}

\author{Timur Bagautdinov}
\email{timurb@meta.com}
\affiliation{%
 \institution{Meta Reality Labs Research}
 \country{USA}
}

\thanks{Work done when DX was a visiting researcher at Meta. ZC and CW are now at Google.}
\renewcommand\shortauthors{Xiang, D. et al}
\renewcommand\shorttitle{Faithful Full-Body Telepresence with Dynamic Clothing Driven by Sparse RGB-D Input}

\begin{abstract}
Clothing is an important part of human appearance but challenging to model in photorealistic avatars. In this work we present avatars with dynamically moving loose clothing that can be faithfully driven by sparse RGB-D inputs as well as body and face motion. We propose a Neural Iterative Closest Point (N-ICP) algorithm that can efficiently track the coarse garment shape given sparse depth input. Given the coarse tracking results, the input RGB-D images are then remapped to texel-aligned features, which are fed into the drivable avatar models to faithfully reconstruct appearance details. We evaluate our method against recent image-driven synthesis baselines, and conduct a comprehensive analysis of the N-ICP algorithm. We demonstrate that our method can generalize to a novel testing environment, while preserving the ability to produce high-fidelity and faithful clothing dynamics and appearance.
\end{abstract}

%
%


%
%


\begin{teaserfigure}
\centering 
\includegraphics[width=\textwidth]{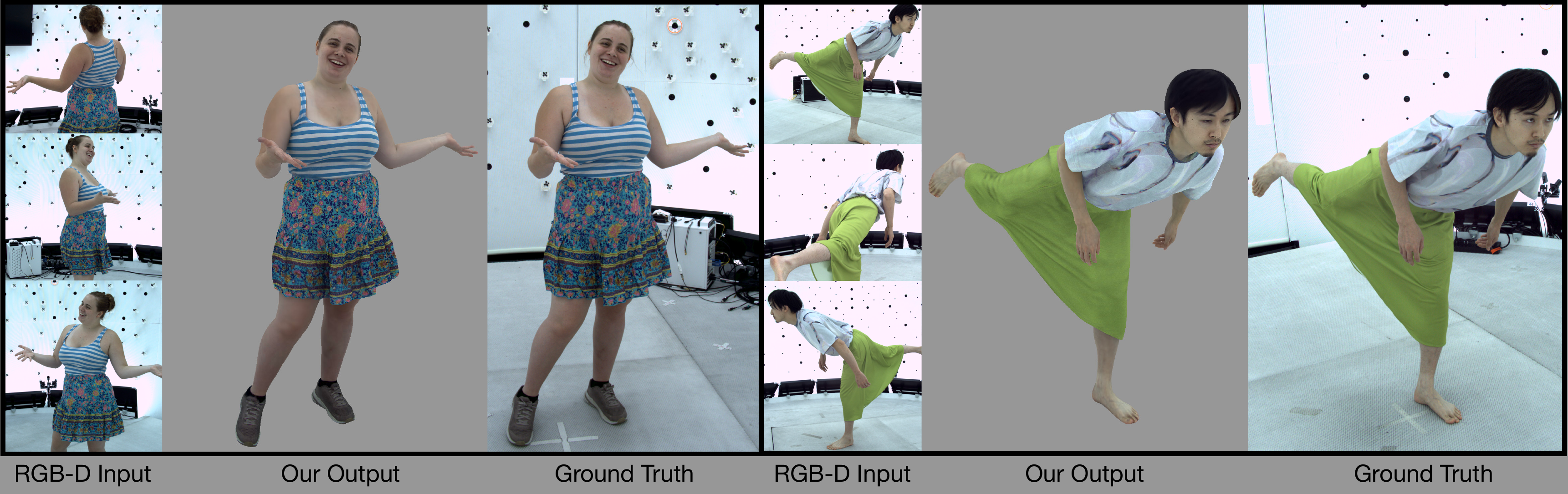} 
\caption{We present photorealistic full-body avatars that can be driven by sparse RGB-D views (along with body pose and facial keypoints) and faithfully reproduce the appearance and dynamics of challenging loose clothing from the input views. We show the input views, our output and the ground truth reference images in each group of results.} 
\label{fig:teaser} 
\end{teaserfigure}

\maketitle

\input{1-introduction.tex}

\input{2-related-work.tex}

\input{3-method.tex}

\input{4-results.tex}

\input{5-conclusion.tex}

\end{document}


\title{Drivable Avatar Clothing: Faithful Full-Body Telepresence with Dynamic Clothing Driven by Sparse RGB-D Input (Supplementary Document)}

\author{Donglai Xiang}
\email{donglaix@cs.cmu.edu}
\affiliation{%
 \institution{Carnegie Mellon University}
 \country{USA}
}

\author{Fabian Prada}
\email{fabianprada@meta.com}

\affiliation{%
 \institution{Meta Reality Labs Research}
 \country{USA}
}

\author{Zhe Cao}
\email{zhecao@berkeley.edu}

\affiliation{%
 \institution{Meta Reality Labs Research}
 \country{USA}
}

\author{Kaiwen Guo}
\email{guokaiwen_neu@126.com}
\affiliation{%
 \institution{Meta Reality Labs Research}
 \country{USA}
}

\author{Chenglei Wu}
\email{chengleiwu@gmail.com}
\affiliation{%
 \institution{Meta Reality Labs Research}
 \country{USA}
}

\author{Jessica Hodgins}
\email{jkh@cmu.edu}
\affiliation{%
 \institution{Carnegie Mellon University}
 \country{USA}
}

\author{Timur Bagautdinov}
\email{timurb@meta.com}
\affiliation{%
 \institution{Meta Reality Labs Research}
 \country{USA}
}

\thanks{Work done when DX was a visiting researcher at Meta. ZC and CW are now at Google.}
\renewcommand\shortauthors{Xiang, D. et al}
\renewcommand\shorttitle{Faithful Full-Body Telepresence with Dynamic Clothing Driven by Sparse RGB-D Input (Supplementary Document)}

\maketitle

\section{Related Work (Continued)}
\emph{Template-Based Performance Capture.} Our work is also related to a group of methods that track human surface by deforming a person-specific or category-specific template or avatar, using either classical optimization \cite{robertini2016model,xu2018monoperfcap,habermann2019livecap,xiang2020MonoClothCap} or network prediction \cite{habermann2020deepcap,habermann2021deeper,jiang2023hifecap,li2022avatarcap}. They achieve better temporal coherency than template-free methods that regress human shape for each frame \cite{saito2019pifu,saito2020pifuhd,li2020monocular,xiu2022icon,xiu2023econ}, but focus on reconstructing human geometry and rather than modeling dynamic appearance.

\section{Ablation Studies on N-ICP}

\begin{table}[t]
\centering
\caption{Ablation studies on different types of input for N-ICP. The evaluation metric is the Mean Squared Error (MSE) in $\si{\milli\metre\squared}$ between surfaces. $\mathbf P, \mathbf r$ and $\mathbf g$ refer to the point cloud, residual and gradient defined in Sec. 4.2. $N$ denotes the number of update iterations. When $N=1$, the network makes a one-shot prediction. Our full method is shown in the last row.}
\begin{tabular}{ L{3.0cm}   C{3.0cm} }
    \toprule
    Input Type & MSE (\si{\milli\metre\squared}) \\
    \midrule
    $\mathbf P ~ (N = 1)$ & 101.19 \\ 
    $\mathbf P, \mathbf r ~ (N = 3)$ & 82.72 \\
    $\mathbf P, \mathbf g ~ (N = 3)$ & 49.60 \\
    $\mathbf P, \mathbf r, \mathbf g ~ (N = 1)$ & 72.30 \\
    $\mathbf P, \mathbf r, \mathbf g ~ (N = 3, \text{full})$ & \textbf{48.47} \\
    \bottomrule
\end{tabular}
\label{table:proposed-eval-nicp-types}
\end{table}

We conduct ablation studies on our design of the N-ICP algorithm. The results are shown in Tab. \ref{table:proposed-eval-nicp-types}. The most naive baseline is to simply use the point cloud as the input feature, shown on the first row of the table. On the second row, we add the closest point residual to the input feature, which provides useful information for surface alignment and enables an iterative update of the deformation parameters. The following rows suggest that the energy gradient derived from the residuals can provide more effective guidance, similar to its critical role in traditional nonlinear optimization. The last two rows verify the benefit of iterative parameter update compared with a one-shot prediction by the network.

\section{Detail of Comparison with Sensing-Based Baselines (Sec. 6.3)}
Here, we provide the implementation detail for the sensing-based baselines for the experiment in Sec. 6.3 in the main paper. We first fuse the sparse input depth maps into a single Truncated Signed Distance Field (TSDF) volume \cite{curless1996volumetric,dong2022ash}, and then extract from it an explicit mesh representation. Using the fused geometry, we can then warp the input RGB images from the source views to any target view. However, the warped image is usually imperfect because the fused geometry is often incomplete and noisy. Therefore, we follow the idea of LookinGood \cite{martin2018lookingood} and train a U-Net to complete the warped image. This baseline essentially learns to inpaint complete human appearance from partial input only in the screen space, and struggles to achieve 3D-aware temporal consistency in the output. As explained in the main paper, this experiment is not intended to be a full-scale comparison against state-of-the-art sensing-based approaches, but to better understand our method in comparison to a modest baseline along this line of work given similar input.

\section{Comparison with Clothing Codec Avatars and Dressing Avatars}

\begin{table}[t]
\centering
\caption{Comparison between Clothing Codec Avatars (CCA) \cite{xiang2021modeling}, Dressing Avatars (DA) \cite{xiang2022dressing} and our method.}
\begin{tabular}{ L{4.5cm}  C{0.8cm}  C{0.8cm}  C{0.8cm}}
    \toprule
     & CCA & DA & Ours \\
    \midrule
    RGB-D driving input & & & \checkmark \\
    Loose clothing dynamics & & \checkmark & \checkmark \\
    Physical simulation & & \checkmark & \\
    Ground truth registration & \checkmark & \checkmark & \\
    Faithful output & & & \checkmark \\
    \bottomrule
\end{tabular}
\label{table:comparison-cca-da-formulation}
\end{table}

We highlight the difference in formluation between our method, Clothing Codec Avatars (CCA) \cite{xiang2021modeling} and Dressing Avatars (DA) \cite{xiang2022dressing} in Tab. \ref{table:comparison-cca-da-formulation}. In terms of driving signal, CCA and DA take body and face motion as input, while our method additionally uses sparse RGB-D views. DA and our method can generate richer and more realistic dynamics for loose clothing than CCA, but DA requires a proprietary implementation of real-time cloth simulation. CCA and DA utilize ground truth clothing registration to train their models, while our method does not require such pre-processing. Finally, thanks to the additional RGB-D input and our model design, our output is more faithful to the actual clothing motion than those two previous methods.

\section{Implementation Detail}

\subsection{Clothing Deformation Graph}

\begin{figure}[t]
    \centering
    \includegraphics[width=\linewidth]{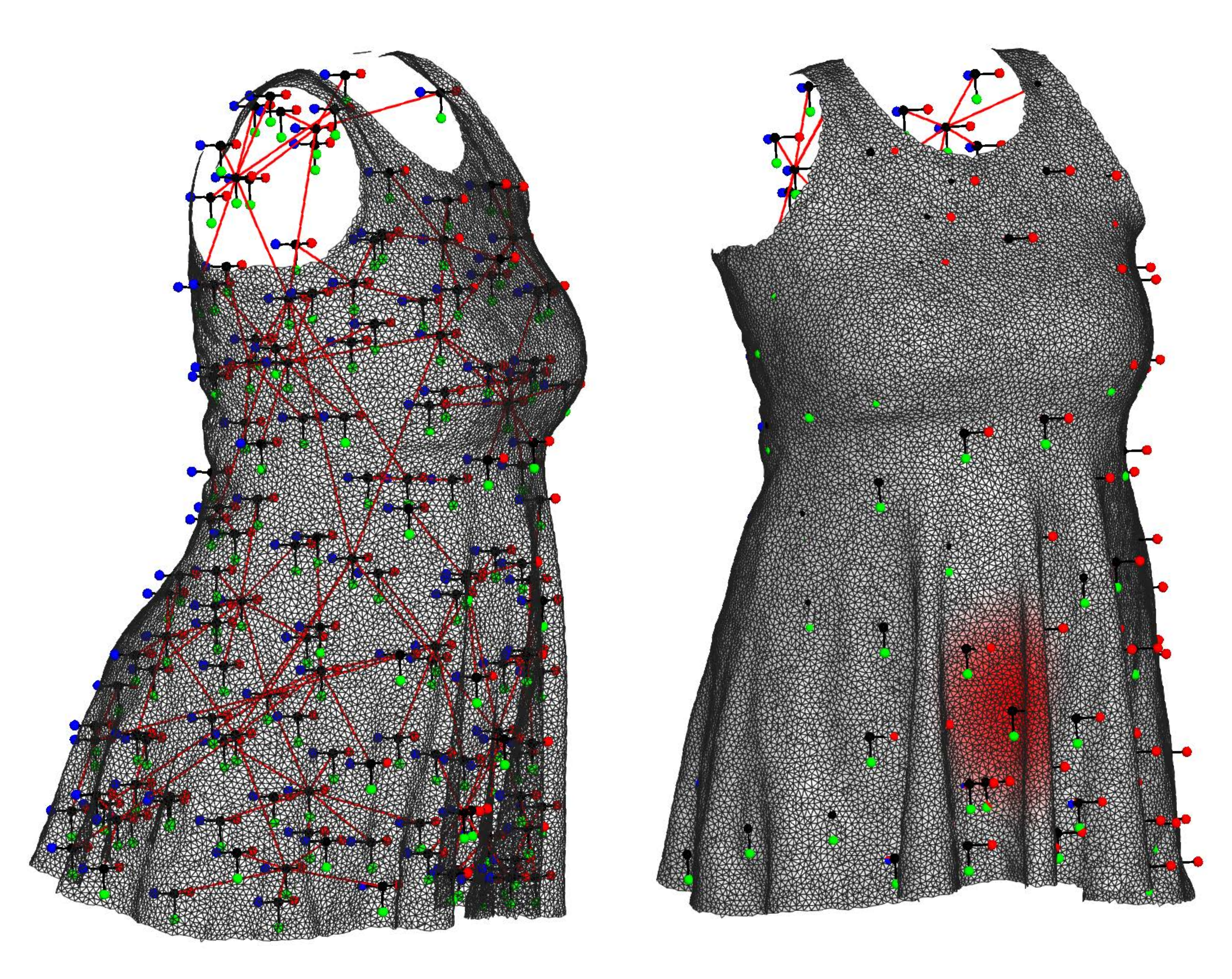}
    \caption{A visualization of the deformation graph $\mathcal E$ used in the dress example. On the left side, we show the coordinate frame at each graph node and their connectivity by the red lines. On the right side, the region of influence by a node located in the center is shown in red.}
    \label{fig:deformation-graph}
\end{figure}

In Fig. \ref{fig:deformation-graph}, we provide a visual illustration of the deformation graph $\mathcal E$ in the inner layer of the clothing deformation model $\mathcal D$ (Sec. 4 of the main paper) for the dress example. The parameters for the deformation graph include the rotation and translation for each node:
\begin{gather}
    \boldsymbol{\theta} = \{\mathbf r_k, \mathbf t_k\}_{k=1}^K, ~ \mathbf r_k, \mathbf t_k \in \mathbb R^3,
\end{gather}
where $\mathbf r_k$ is the axis-angle representation of a 3D rotation. We use a total of $K = 125$ nodes for each example.

\subsection{Training Setup}

\subsubsection{N-ICP}

When training the N-ICP module, we adopt a regularization term for deformation graph that compares the difference in transformation between adjacent nodes:
\begin{gather}
    L_\text{DG-Reg} = \frac 1{K(K-1)} \sum_{1 \le j \ne k \le K} \Vert T_j \mathbf m_{jk} - T_k \mathbf m_{jk} \Vert^2,
\end{gather}
where $T_j$ and $T_k$ denote the SE(3) transformation for the $j-$th and $k-$th nodes respectively, and $\mathbf m_{ij}$ denotes the middle point between the rest positions of the $j-$th and $k-$th nodes. Then the total loss function for training N-ICP is written as
\begin{gather}
    L_\text{N-ICP} = \frac1N \sum_{i=1}^N L_\text{ICP}(\boldsymbol{\theta}^{(i)}, \overline{\mathbf P}) + \lambda_\text{DG-Reg} L_\text{DG-Reg},
\end{gather}
where the balancing weight is set to $\lambda_\text{DG-Reg} = 1 \times 10^{-3}$. The trainable parameters in N-ICP are those in PointNet $\mathcal M$. The input and output of the PointNet $\mathcal M$ are converted to the root body coordinate of the subject given the tracked body pose $\boldsymbol \rho$ to be invariant to the global orientation and translation. We use the AdamW optimizer with an initial learning rate of $1 \times 10^{-5}$.

\emph{Initialization.} We find it crucial to initialize the parameters in the last layer of the PointNet with values close to zero, so that $\boldsymbol{\theta}^{(i)} \approx \mathbf{0}$ for $i = 1, \dots, N$ at the first training iteration, with $\boldsymbol{\theta}^{(0)}$ set to $\mathbf{0}$. In this way, thanks to the two-layer clothing deformation model (Sec. 4 in the main paper), $\mathcal D(\boldsymbol{\theta}^{(i)})$ is close enough to the ICP target to generate meaningful gradient at the beginning of the training process, and gradually converges to the desired minimum. In practice, we initialize the last layer of the network by random sampling from a uniform distribution $U[-\varepsilon, \varepsilon]$ where $\varepsilon = 1 \times 10^{-6}$.

\emph{Discussion on supervision.} N-ICP is trained in a self-supervised manner, because the loss function $L_\text{N-ICP}$ does not involve the ``ground truth'' deformation parameters. The reasons are two-fold. First, it takes extra processing time efforts obtain the ground truth. Second, the problem of estimating reliable “ground truth” deformation parameters is challenging by itself. Unless the garment under capture has been specially designed to encode correspondences in a printed pattern \cite{halimi2022garment}, otherwise, the principal approach is to run offline ICP between the deformation model and MVS geometry. In this way, the ``ground truth'' essentially offers no more information than directly supervising N-ICP by MVS. The self-supervised formulation, instead, allows solving a global optimization by sharing the information across all frames.

\subsubsection{Texel-Conditioned Clothed Avatars}

We use the following loss functions to train the texel-conditioned clothed avatars (Sec. 5 of the main paper)
\begin{gather}
    L_\text{avatars} = \sum_{i}\lambda_\text{i} L_\text{i}, ~ i \in \{\text{RGB}, \text{mask}, \text{reg}, \text{part}, \text{ID-MRF}\}.
\end{gather}
$L_\text{RGB}$ and $L_\text{mask}$ are the standard $L_1$ losses for RGB and mask respectively; $L_\text{reg}$ is the Laplacian regularization terms for body and clothing meshes. $L_\text{part}$ is similar to $L_\text{mask}$ but identify background, body and clothing in three different categories. Following \cite{feng2022capturing}, we use the ID-MRF loss \cite{wang2018image}, a stronger form of perceptual loss to encourage sharpness for high-frequency texture in the clothing region. We use $\lambda_\text{RGB} = 0.2, \lambda_\text{mask} = \lambda_\text{part} = 500.0, \lambda_\text{reg} = 100.0, \lambda_\text{ID-MRF} = 1.0$. The gradient of loss functions defined in the image space (RGB, mask, part and ID-MRF) with respect to the network parameters are back-propagated through a differentiable rasterizer. We use the AdamW optimizer with a learning rate of $1 \times 10^{-3}$.

\emph{Color Augmentation.} In order to deal with the domain gap in illumination and color when the directly applying the avatars to the novel capture environment (Sec. 6.4 in the main paper), we apply a random color augmentation to texel-aligned RGB features $\mathbf F_I$ using the `ColorJitter' function in TorchVision\footnote{\url{https://github.com/pytorch/vision}. We use the following parameters: brightness=0.5, contrast=0.5, saturation=0.5, hue=0.2.} at training time. Notice that we leave the ground truth images used for supervision in $L_\text{avatars}$ unchanged, so that the network always preserves the original appearance in the \textit{output}, despite a different color mode in the \textit{input} feature $\mathbf F_I$ when we direct apply the model to the novel environment. The output appearance only changes after fine-tuning with ground truth images in the novel environment.

\begin{figure*}[t]
    \centering
    \includegraphics[width=\linewidth]{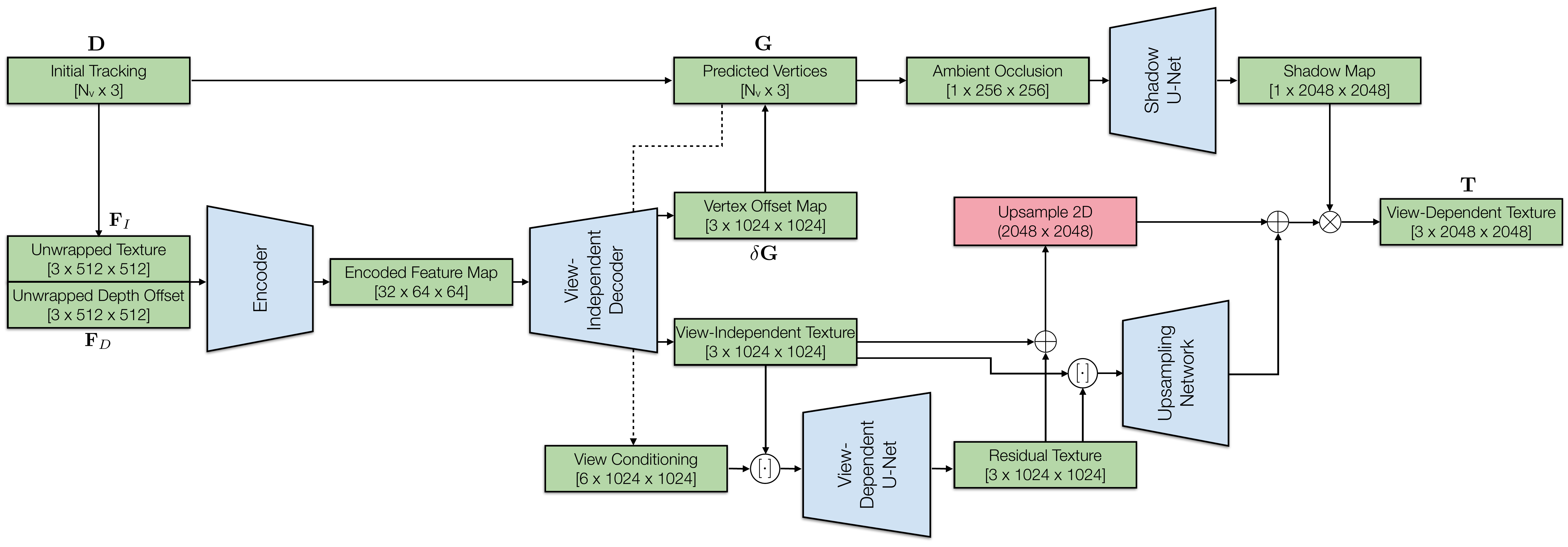}
    \caption{The network architecture for the texel-conditioned clothed avatars. It consists of the following five components: (1) a convolutional encoder that encodes the texel-aligned input, (2) a view-independent decoder that outputs vertex and texture maps, (3) a view-dependent U-Net that regresses view-dependent variation in the texture, (4) a shadow network that takes in ambient occlusion and computes a multiplicative shadow map, and (5) an upsampling network that predicts the residual after increasing the spatial resolution from 1024 to 2048.}
    \label{fig:avatar_architecture}
\end{figure*}

\subsection{Preprocessing and Postprocessing}

\subsubsection{Input Preprocessing.} Our method takes RGB and depth images as input. When training and testing using data from the dense capture studio, we run image-based part segmentation and transfer the result to the MVS mesh by projection and visibility check. This operation allows us to extract the clothing region. The MVS mesh may include floating noise, which we remove by checking the mesh connectivity and setting a threshold on the minimal number of vertices in a connected component. Then, we rasterize the segmented mesh to RGB views to ``simulate'' a depth image.

When training and testing in the novel environment, we use the RGB-D images from calibrated Kinect sensors as input. We also run image-based part segmentation to extract the clothing regions. Then we use TSDF fusion \cite{curless1996volumetric} and Marching Cubes \cite{lorensen1998marching} to form a mesh from the extracted depth images, which allows us to perform similar connectivity check as above to remove noise from the depth sensors. 

\subsubsection{Temporal Smoothing.} Due to the unstructured point cloud input, the output of N-ICP may have undesirable jittering. We apply temporal smoothing to the output of N-ICP by taking the average on the vertex positions in a small temporal window, which is feasiable because the N-ICP output shares a consistent registered topology across all the frames. The filtered meshes are then used to unwrap texel-aligned features and as input to the texel-conditioned avatars as shown in Fig. 2 of the main paper. We find no need to apply additional smoothing on the final output of texel-conditioned avatars if the provided initial tracking is temporally stable and the floating depth noise has been removed in the preprocessing step.

\subsubsection{Collision.} To resolve the collision between the body and clothing layers, which is usually slight in the results, we follow Clothing Codec Avatars \cite{xiang2021modeling} (Sec. 6) to project the clothing vertices in collision beyond the nearest body points by a slight margin. More sophisticated ways to handle collision based on geometry or learning \cite{tan2022repulsive} may be incorporated, which we leave for future work.

\subsection{Network Architecture}

\subsubsection{N-ICP} N-ICP takes an unstructured point cloud as input, so we adopt the PointNet++ \cite{qi2017pointnet++} architecture. To specify the architecture, we reuse the notation of Set Abstraction function from \cite{qi2017pointnet++}:
$$ \text{SA}(K, r, [l_1, \dots, l_d]), $$
where $K$ denotes the number of grouping centers, $r$ denotes the radius of the grouping regions, and $l_i$ denotes the output size of a fully connected layer in the Multi-Layer Perceptron (MLP). We also denote a standalone MLP as $\text{FC}([l_1, \dots, l_d])$. Then the architecture of the network $\mathcal M$ can be described as
\begin{gather*}
[\mathbf p, \mathbf r] \rightarrow
\text{SA}(32, 0.1, [16, 16, 32]) \rightarrow
\text{SA}(32, 0.2, [64, 64, 128]) \rightarrow \\
\text{SA}(32, 0.4, [256, 256, 256]) \rightarrow 
\text{SA}(32, 0.8, [256, 256, 512]) \rightarrow \\
\text{MaxPool} \rightarrow \bigoplus \mathbf g \rightarrow \text{FC}([512, 512, 512, 750]) \rightarrow \Delta \boldsymbol{\theta},
\end{gather*}
where $\mathbf p$ and $\mathbf r$ denote the point coordinate and residual as defined in Sec. 4 of the main paper, and $\bigoplus \mathbf g$ denotes the operation to concatenate the result from the previous step with gradient input $\mathbf g$.

\subsubsection{Texel-Conditioned Clothed Avatars} The overall architecture of the texel-conditioned avatar models is shown in Fig. \ref{fig:avatar_architecture}. Given the texel-aligned features $\mathbf F_I, \mathbf F_D$ unwrapped from the initial tracking results $\mathbf D$ as input, the encoder produces a feature map that is spatially aligned with the input. The encoded feature maps are then decoder into a vertex offset map $\delta \mathbf G$, from which the offsets are extracted and then applied on top of the initial tracking to obtain the output geometry $\mathbf G$. The geometry $\mathbf G$ and the viewpoint $\mathbf v$ are used together to compute the view-conditioning, including the viewing vector expressed in the local Tangent-Bitangent-Normal (TBN) frame \cite{xiang2022dressing} as well as its reflected direction. The view-dependent U-Net takes in the view conditioning and the view-independent texture to produce an additive view-dependent offset. With the final geometry $\mathbf G$, we also compute the ambient occlusion, which is fed into the shadow U-Net to produce a multiplicative shadow map. The view-dependent texture is then upsampled to 2k resolution by a upsampling network. 

\begin{figure*}[t]
    \centering
    \includegraphics[width=\linewidth]{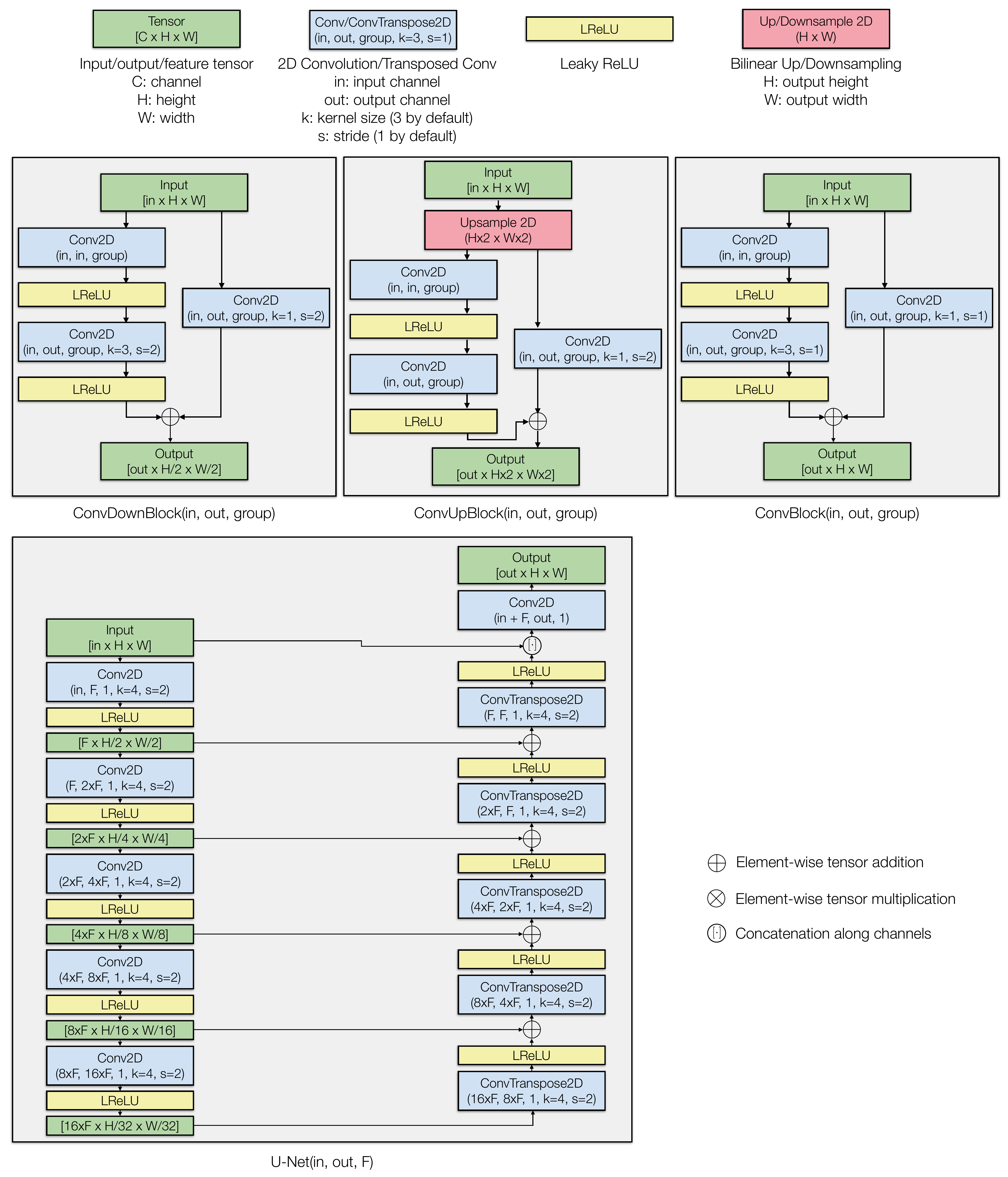}
    \caption{Network blocks used in the architecture of texel-aligned avatars.}
    \label{fig:legend}
\end{figure*}

To specify the architecture of the individual networks above, we define the blocks shown in Fig. \ref{fig:legend}.

\emph{(1) Convolutional encoder} consists of the network blocks in the following table. Following DVA \cite{remelli2022drivable}, we find that using a U-Net at $64 \times 64$ resolution instead of a bottleneck structure helps to preserve the UV-space detail in the output.

\begin{table}[h]
\centering
\begin{tabular}{c c}
    \toprule
    Block & Output Size ($C \times H \times W$) \\
    \midrule
    ConvBlock(6, 16, 1) & $16 \times 512 \times 512$ \\
    ConvDownBlock(16, 32, 1) & $32 \times 256 \times 256$ \\
    ConvDownBlock(32, 64, 1) & $64 \times 128 \times 128$ \\
    ConvDownBlock(64, 64, 1) & $64 \times 64 \times 64 $ \\
    U-Net(64, 64, 32) & $32 \times 64 \times 64 $ \\
    \bottomrule
\end{tabular}
\end{table}

\emph{(2) View-independent decoder} consists of the network blocks in the following table. Here, the ``RepeatChannels'' operation repeats the channels of the input feature for the geometry and texture branches. The following ``ConvUpBlocks'' processing them separately in different groups. The output is then evenly split into a vertex offset map and a texture map.

\begin{table}[h]
\centering
\begin{tabular}{c c}
    \toprule
    Block & Output Size ($C \times H \times W$) \\
    \midrule
    ConvBlock(32, 32, 1) & $32 \times 64 \times 64$ \\
    RepeatChannels & $64 \times 64 \times 64$ \\
    ConvUpBlock(64, 32, 2) & $32 \times 128 \times 128$ \\
    ConvUpBlock(32, 16, 2) & $16 \times 256 \times 256$ \\
    ConvUpBlock(16, 8, 2) & $8 \times 512 \times 512$ \\
    ConvUpBlock(8, 8, 2) & $8 \times 1024 \times 1024$ \\
    Conv2D(8, 6, 2, k=1) & $6 \times 1024 \times 1024$ \\
    SplitChannels & $(2 ~ \times) ~ 3 \times 1024 \times 1024$ \\
    \bottomrule
\end{tabular}
\end{table}

\emph{(3) View-dependent U-Net} is a single block ``U-Net(9, 4, 3)'' defined in Fig. \ref{fig:legend}.

\emph{(4) Shadow U-Net} is an upsampling operation on the input ambient occlusion map from 256 resolution to 2048, followed by a block ``U-Net(1, 2, 1)''.

\emph{(5) Upsampling network} is defined in the following table. Here the ``PixelShuffle$(r)$'' is an operation that rearranges a tensor from shape $(C \times r^2) \times H \times W$ to $C \times (H \times r) \times (W \times r)$.

\begin{table}[h]
\centering
\begin{tabular}[h]{c c}
    \toprule
    Block & Output Size ($C \times H \times W$) \\
    \midrule
    Conv2D(6, 2, 1) & $2 \times 1024 \times 1024$ \\
    LReLU(0.2) & $2 \times 1024 \times 1024$ \\
    Conv2D(2, 12, 1) & $12 \times 1024 \times 1024$ \\
    PixelShuffle(2) & $3 \times 2048 \times 2048 $ \\
    \bottomrule
\end{tabular}
\end{table}

\subsection{Training Data Preparation}

In this section, we describe how we prepare the assets required to train the avatars. Given the multi-view images captured by more than one hundred synchronized cameras, we run 2D keypoint detection, part segmentation, and Multi-View Stereo (MVS). The 2D body keypoints are triangulated to estimate 3D keypoints. For each vertex in the MVS output, we aggregate its category label from each camera view by checking its projection in the image segmentation, and then perform a majority voting, followed by a Markov Random Field (MRF) to ensure spatial smoothness. We also use the method in \cite{zhang2017detailed} to estimate an underlying body template and the body pose for each frame given the 3D keypoints and segmented MVS mesh. The whole process is similar to \cite{xiang2021modeling}, except that we do not perform clothing registration offline in the style of ClothCap \cite{pons2017clothcap}. Instead, we define a deformation model $\mathcal D(\boldsymbol \theta)$, and train the N-ICP network to track the clothing in a self-supervised manner. The clothing template is created from the segmented clothing region in the MVS mesh in a rest-pose frame with some manual cleanup and remeshing.

\bibliographystyle{ACM-Reference-Format}
\bibliography{local}

%% file: 1-introduction.tex
\section{Introduction}

Photorealistic avatars are important for enabling truly immersive and believable telepresence experiences.
%
An ideal telepresence application should not only produce plausible-looking results, 
but also be complete and accurate: all salient aspects of human appearance should be 
captured and resynthesized to fully match the real-world states of the subject.
However, these properties are particularly challenging to achieve with clothing, 
an integral part of human appearance,
due to its complex movements on human body. 

To deal with this challenge, some previous methods go beyond treating
clothing effectively as a part of the human body~\cite{bagautdinov2021driving,liu2021neural}
and perform explicit modeling of clothing as a separate layer on top of the underlying body~\cite{xiang2021modeling,xiang2022dressing}. 
These methods can work well for pose-driven animation, i.e., synthesizing plausible clothing deformation 
and photorealistic appearance that are perceptually compatible with the input pose signal. 
However, there is no guarantee that the animation output will faithfully reproduce the actual 
states of clothing (Fig.~\ref{fig:comparison-prev-cloth-work}), 
and potentially distorting the conveyed social signals. 
In fact, the dynamics of clothing cannot be fully explained by the body pose of the current frame or
a few previous frames. Given two distinct initial states of clothing, the same body motion can result in 
completely different trajectories of clothing deformations, especially for loose garments like skirts or 
dresses. 
Therefore, it is impossible to infer accurate clothing states given such incomplete input signals.

An alternate approach for telepresence relies heavily on the availability of sensory inputs without a strong human prior. 
For example, volumetric fusion methods~\cite{newcombe2015dynamicfusion,dou2016fusion4d,dou2017motion2fusion} 
produce a complete geometrical representation of a scene by tracking and fusing observations from sparse 
RGB-D cameras. 
Neural implicit functions can also be utilized to reconstruct a dynamically moving human surface from sparse
camera inputs~\cite{yu2021function4d,shao2022floren}, or even to directly model the radiance field of clothed
human appearance~\cite{lin2022efficient,shao2022doublefield}. 
In theory, these methods are flexible enough to be able to reconstruct arbitrary shape from the given input streams. 
However, due to a lack of model constraints, it is generally more challenging for these methods to achieve high-fidelity 
temporal coherency especially with noisy or incomplete input, and the output quality is heavily tied with the sensory input.
For example, it is hard to produce the sharp and subtle detail in hands \cite{shao2022doublefield} with the limited resolution and noise, and the observed and unobserved regions from the input can have obvious difference in the output quality \cite{shao2022floren}.
Human priors have been introduced to regularize the predictions~\cite{kwon2021neural,kwon2023neural}, but the ability to reliably handle loose clothing has not been clearly demonstrated.

To leverage the benefits of both families of approaches, we can rely on explicit 
avatar models as a prior, but expand the driving signal to include the denser input in addition to the body pose.
We build avatars with dynamic clothing that can be driven from a sparse set of RGB-D cameras (usually three unless otherwise stated).
This formulation allows for more faithful resynthesis of the human appearance, including clothing details.
%
We build on top of DVA~\cite{remelli2022drivable}, which proposes the texel-conditioned avatar, an encoder-decoder
model that takes in UV-aligned driving features and predicts geometry and appearance for rendering. However, DVA only works
well for tight clothing that closely follows underlying body, due to the limitation in relying on a generic body shape prior. 

To better handle loose clothing, our insight is to introduce a tracking stage that 
coarsely aligns the loose clothing surface with the input depth. 
More specifically, we propose a simple-yet-effective Neural Iterative Closest Points (N-ICP) algorithm 
to iteratively update a clothing deformation model given the feedback from surface error 
in a data-driven manner. 
N-ICP combines the flexibility of the classical ICP methods which allows
us to handle large clothing deformations, while relying on learning for more
efficient inference and reliable geometry estimates.
In contrast to DVA, which uses coarse body geometry to extract features, the N-ICP tracking allows us to extract 
more accurate and meaningful texel-aligned features. It also eases
the burden on the encoder-decoder model, since large deformations and misalignments 
are handled by the coarse tracking, and ultimately leads to better quality and generalization.
%
%
In addition, several technical components have been leveraged to further improve the texel-conditioned avatars.
To aid geometry prediction, we augment texel-aligned features with geometry 
features computed from depth and coarsely tracked geometry. 
To improve appearance, we adopt a specific perceptual loss to encourage 
high-frequency texture detail on the predicted clothing.

Our contributions can be summarized as follows:
\begin{itemize}
    \item We develop photorealistic full-body avatars with dynamic clothing that faithfully reproduces the original states of subject's appearance and geometry. 
    These avatars are driven by sparse (up to 3) RGB-D inputs and enable free-viewpoint rendering.
    \item As an important component of our system, we introduce a Neural Iterative Closest Point algorithm that learns to iteratively find the most effective parameter update 
    to track the input point cloud efficiently with a deformation model.
    \item Our avatars can directly generalize to a novel testing environment with a different background and illumination, while capturing the complex clothing dynamics and preserving the original high-quality appearance from the training data. We also provide an option to finetune the model to adapt to the new appearance in the novel environment.
\end{itemize}

%% file: 2-related-work.tex
\begin{figure*}[t]
    \centering
    \includegraphics[width=0.95\linewidth]{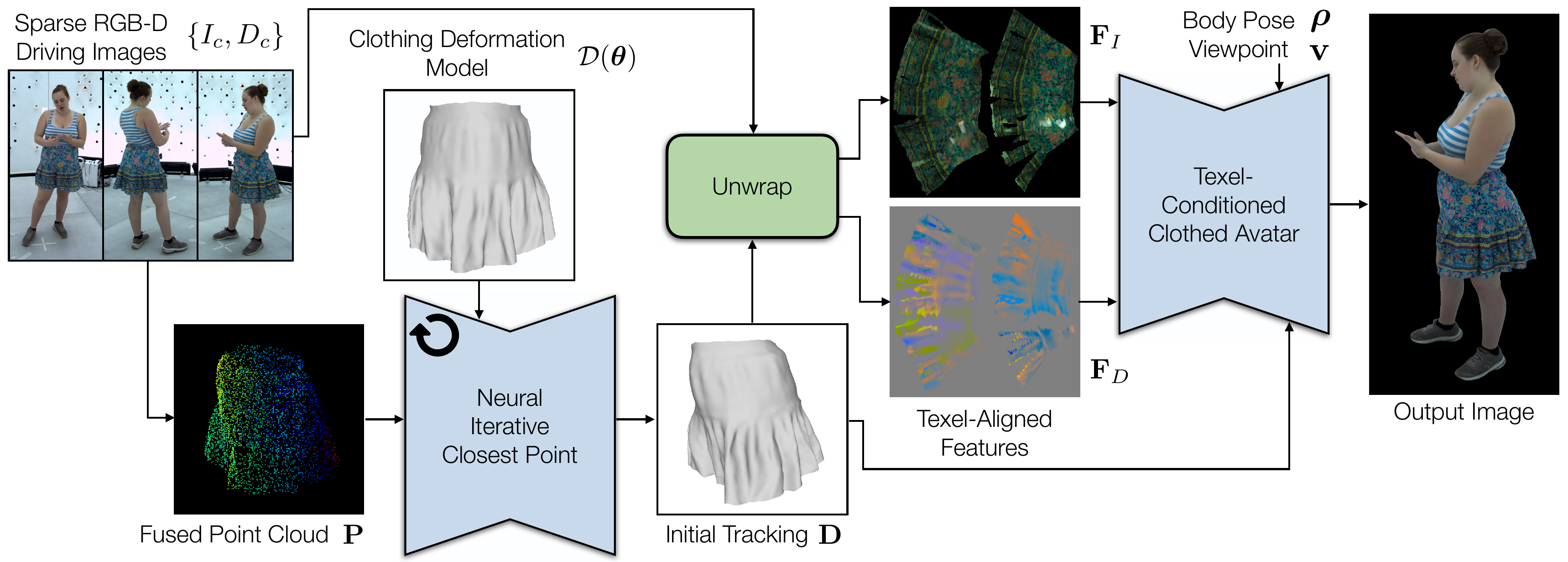}
    \caption{An overview of our method. The Neural Iterative Closest Point module efficiently tracks the clothing surface from the input point cloud $\mathbf P$ with a clothing deformation model $\mathcal D(\boldsymbol\theta)$; the initial tracking result $\mathbf D$ is then used to unwrap the driving images $I_c$ and depth maps $D_c$ into texel-aligned features $\mathbf F_I, \mathbf F_D$, which are then fed into the texel-conditioned avatar, together with body pose $\boldsymbol{\rho}$, facial keypoints and viewpoint $\mathbf v$, to predict the output image.}
    \label{fig:proposed-method-overview}
\end{figure*}

\section{Related Work}




We focus on related work in three areas: photorealistic clothed avatars, sensing-based telepresence, and image-conditioned novel-view synthesis and learning to optimize for non-rigid tracking.

\paragraph{Photorealistic Clothed Avatars.} Here we discuss approaches pose-driven avatars that model both the shape and photorealistic appearance of clothed humans leanrned directly from captured images in a data-driven manner. Depending on the appearance model, these methods can be roughly divided into three categories: methods that combine coarse human geometry with Deferred Neural Rendering \cite{thies2019deferred,grigorev2021stylepeople,raj2021anr}, methods that incorporate a human prior into Neural Radiance Fields \cite{mildenhall2020nerf,peng2021animatable,liu2021neural,li2022tava,su2021nerf,zheng2022structured}, and methods based on mesh and dynamic texture \cite{shysheya2019textured,habermann2021real}. Our method is more closely related to the Full-Body Codec Avatars \cite{bagautdinov2021driving} among the third category, as well as the followup work that can handle dynamic clothing \cite{xiang2021modeling,xiang2022dressing}. Most of these methods learn a mapping from body pose or motion to clothed body appearance, without resolving the ambiguity in clothing states for similar body poses. Drivable Volumetric Avatars (DVA) \cite{remelli2022drivable} is an exception, which is additionally conditioned on a dynamic texture unwrapped from sparse driving views to incorporate the clothing information. However, it is limited to tight clothing due to a constraint in body representation. Our method can further track and render dynamically moving loose clothing realistically.


\paragraph{Sensing-Based Telepresence Approaches.} Another category of approaches for telepresence rely more heavily on the sensing input for surface reconstruction, usually from one or several RGB(-D) cameras. Volumetric fusion  \cite{newcombe2015dynamicfusion,dou2016fusion4d,dou2017motion2fusion} and multiview-conditioned implicit functions \cite{suo2021neuralhumanfvv,yu2021function4d,shao2022floren} have been utilized to reconstruct scene geometry from the sensor input. Early work only reconstructs the geometry \cite{newcombe2015dynamicfusion}, but later work also predicts color by warping and blending RGB information from input views \cite{dou2016fusion4d,lawrence2021project}, possibly assisted by deep neural networks \cite{suo2021neuralhumanfvv,yu2021function4d,shao2022floren}. Neural rendering has also been introduced \cite{martin2018lookingood,nguyen2022free} to compensate for artifacts in the reconstructed geometry. These sensing-based methods enjoy the flexibility in handling varying topology, but are generally more sensitive to noisy or missing input than the model-based approaches described in the previous section.

\paragraph{Image-Conditioned Novel-View Synthesis.} Our task can also be regarded as Novel-view Synthesis (NVS) based on sparse input images. Generalizable NeRF is a group of methods that extend the Neural Radiance Field (NeRF) \cite{mildenhall2020nerf} and allow the reconstruction of a scene with sparse images as input without per-scene optimization \cite{yu2021pixelnerf,wang2021ibrnet,chen2021mvsnerf,lin2022efficient,shao2022doublefield}. This formulation is thus more suitable for telepresence than the original NeRF, but tends to perform less well when the target view is far away from the sparse input views due to the inherent 3D ambiguity. To alleviate this problem, some methods incorporate prior knowledge of the human body to achieve better quality. Neural Body \cite{peng2021neural} utilizes the SMPL model to aggregate the temporal information over the multiview videos. It still does not allow direct use of novel images as input, but this limitation is addressed in Neural Human Performer \cite{kwon2021neural} and GP-NeRF \cite{chen2022geometry}. Several other methods \cite{zhao2022humannerf,gao2022mps,gao2023neural,kwon2023neural} predict radiance fields in a canonical space with the help of forward and inverse skinning transformations. These methods additionally allow generalization across identities, which is not the focus of our work. KeypointNeRF \cite{mihajlovic2022keypointnerf} utilizes 3D body keypoints to encode the spatial information in the rendered volume. These methods have shown limited capability to handle dynamic loose clothing due to the body representation.

\paragraph{Learning to Optimize for Non-Rigid Tracking.} Many non-rigid tracking and reconstruction problems are traditionally formulated as optimization. Examples include SMPL model fitting for 3D human pose estimation \cite{bogo2016keep}, and non-rigid ICP for deformable surface tracking \cite{guo2015robust}.
In recent years, researchers attempted to incorporate deep neural networks into these problems \cite{bhatnagar2020loopreg,bozic2021neural,li2020learning,bozic2020neural}.
Our N-ICP formulation essentially treats the neural network as an optimization solver that iteratively generates a parameter update. Some work \cite{song2020human,corona2022learned} explores a similar idea for monocular human pose reconstruction in a supervised setting. RMA-Net \cite{feng2021recurrent} also addresses the non-rigid registration problem with recurrent update. In addition to the difference in deformation model and loss function, our method integrates learning into optimization more deeply by explicit feeding the error and gradient into the network for update prediction, which is shown to be the key to effectiveness in the ablation study.

%% file: 3-method.tex
\section{Method Overview}

Our method takes as input RGB-D images from sparse (up to 3) views, as well as 3D body pose in the form of joint angles (and facial keypoints if available), and generates photorealistic rendering of the subject from an arbitrary viewpoint. 
The model is trained on images of the subject captured in a dense camera system. 
We adopt the two-layer representation that has proven effective in previous work on loose clothing~\cite{xiang2021modeling,xiang2022dressing}.
Our method consists of two major modules. First, in the Neural Iterative Closest Point module, we coarsely track the loose clothing surface given the fused point cloud
from the input depth maps using a deformation graph model. 
Second, we convert the sparse driving RGB-D images into texel-aligned features and feed them into texel-conditioned clothed avatars to infer detailed geometry and view-dependent texture, which are then rasterized to form the output image. 
The overall pipeline is illustrated in Fig. \ref{fig:proposed-method-overview}.


\section{Neural Iterative Closest Point}
\label{sec:proposed-nicp}

As the first step of our approach, we introduce a Neural Iterative Closest Point (N-ICP) algorithm 
to coarsely track the dynamic clothing surface using a deformation graph representation 
of the clothing geometry. Such a module is needed for two reasons. 
First, coarse tracking provides rough correspondences on the clothing surface across different frames. 
Previous work \cite{xiang2021modeling,xiang2022dressing} shows that such canonicalization reduces the variance 
in appearance that needs to be modelled by the downstream module and leads to improved quality. 
Second, compared with skeleton-level tracking~\cite{remelli2022drivable}, the deformation graph 
enjoys the flexibility to track the surface at a higher accuracy from the input depth, 
so that the following stage (Sec.~\ref{sec:proposed-dense-avatar}) only needs to predict a small geometry correction. 
This concept of coarse-to-fine modeling has also proven useful in previous work, e.g.~\cite{habermann2021real}.

\paragraph{Non-Rigid ICP} The classical approach to track a deforming surface is 
the non-rigid Iterative Closest Point (ICP) algorithm~\cite{li2008global}. 
Given a deformation model $\mathcal D$ controlled by deformation parameters $\boldsymbol \theta$, 
the non-rigid ICP algorithm tracks an input point cloud\footnote{At test time, the point cloud comes from fused driving depth images. We only use the points from the regions of dynamic clothing identified by image segmentation.} $\mathbf P$ by solving the following minimization problem
\begin{gather}
    \min_{\boldsymbol\theta} L_\text{ICP}(\boldsymbol\theta; \mathbf P) = \min_{\boldsymbol\theta} \sum_{\mathbf p \in \mathbf P} \Vert \mathcal C(\mathbf p, \mathcal D(\boldsymbol\theta)) - \mathbf p \Vert^2,
    \label{eq:proposed-loss-icp}
\end{gather}
where $\mathcal C$ queries the closest point on the deformed mesh $\mathcal D(\boldsymbol\theta)$ from a point $\mathbf p$ in the point cloud based on Euclidean or projective distance. To keep the deformed mesh in a smooth shape, some regularization terms are often used together with Eq.~(\ref{eq:proposed-loss-icp}) to penalize extreme distortion. This non-linear optimization problem is usually solved by the Gauss-Newton method, especially its Levenberg–Marquardt variant. Concretely, in each iteration of optimization, the algorithm evaluates the residual vector $\mathbf r$, which consists of offset $\mathbf r(\mathbf p) = \mathcal C(\mathbf p, \mathcal D(\boldsymbol\theta)) - \mathbf p$ for each point $\mathbf p$ in the point cloud, and the Jacobian $\mathbf J$ of each residual term with respect to the parameters. Then the algorithm searches for the update $\Delta\boldsymbol\theta$ by solving the linear system $(\mathbf J^T \mathbf J + \lambda \mathbf I) \Delta \boldsymbol\theta = - \mathbf J^T \mathbf r$, where $\lambda > 0$ is a damping coefficient. The parameters are updated by $\boldsymbol\theta^{(i+1)} = \boldsymbol\theta^{(i)} + \Delta\boldsymbol\theta$, where $i$ denotes the index of iteration.

\begin{figure}[t]
    \centering
    \includegraphics[width=\linewidth]{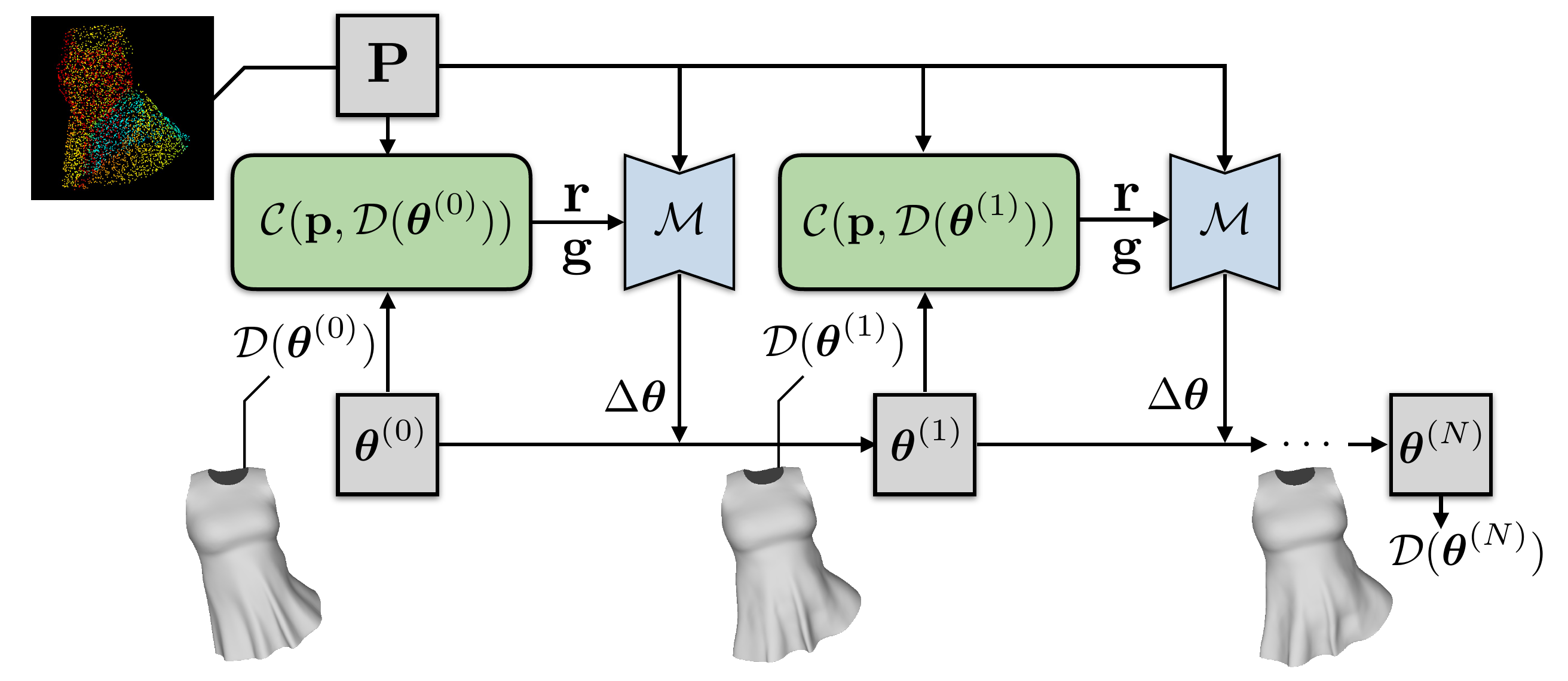}
    \caption{The Neural Iterative Closest Point algorithm. In each iteration, we perform the closest point query between the input point cloud $\mathbf P$ and deformed clothing model $\mathcal D(\boldsymbol\theta^{(i)})$. The residual $\mathbf r$ and gradient $\mathbf g$ are then passed to a neural network $\mathcal M$, which finds the best update $\Delta\boldsymbol\theta$ to the parameters. This process is repeated for $N$ iterations.}
    \label{fig:proposed-nicp}
\end{figure}

\paragraph{N-ICP} The problem of driving avatars for online telepresence poses a challenge in terms of both robustness and speed for the non-rigid tracking algorithm. The nonlinear minimization problem in Eq. (\ref{eq:proposed-loss-icp}) relies heavily on good initialization, so classical method usually requires sequential tracking that is hard to recover from failure and technically demanding GPU implementation to meet the runtime constraint \cite{zollhofer2014real,newcombe2015dynamicfusion}.

This challenge motivates us to introduce a Neural Iterative Closest Point (N-ICP) algorithm, where the goal is to leverage the prior learned by a deep neural network to make an efficient and robust prediction of the update direction $\Delta\boldsymbol{\theta}$. We utilize a PointNet \cite{qi2017pointnet++} architecture to operate on the input point cloud $\mathbf P$. In addition to the coordinates of each point $\mathbf p$, we also use the offset $\mathbf r(\mathbf p) = \mathcal C(\mathbf p, \mathcal D(\boldsymbol\theta)) - \mathbf p$ as a feature, which includes the essential closest point information for solving the non-rigid ICP problem. Inspired by the classical optimization paradigm, we also a provide first-order derivative to the network. For the ease of computation, we use the gradient $\mathbf g = \mathbf J^T \mathbf r$ of total loss with respect to the parameters, which can be automatically derived in most modern neural network libraries \cite{paszke2019pytorch}, or manually derived for better computation efficiency. The formulation of the network $\mathcal M$ can be written as $\Delta \boldsymbol \theta = \mathcal M(\mathbf P, \mathbf r, \mathbf g)$.
The update $\boldsymbol\theta^{(i+1)} = \boldsymbol\theta^{(i)} + \Delta\boldsymbol\theta$ is then performed for $N = 3$ times.

\paragraph{Weakly-Supervised Learning} We train the network in a \textit{weakly-supervised} manner without requiring the ground truth deformation parameters $\boldsymbol \theta$. Instead, we only use the clothing geometry $\overline{\mathbf P}$ reconstructed by Multi-View Stereo (MVS). Compared with $\mathbf P$ which has missing parts due to occlusion from the sparse depth input, $\overline{\mathbf P}$ includes the complete clothing geometry reconstructed from all cameras in the capture studio, and is only used at training time for supervision. Our loss function for training the network is written as $L_{\text{N-ICP}} = \frac1N \sum_{i=1}^N L_\text{ICP} (\boldsymbol \theta^{(i)}, \overline{\mathbf P})$ plus regularization (see supplementary document). In other words, the network is trained to find an update to the parameters so that the deformed mesh can most efficiently track the clothing surface for every ICP iteration.

\paragraph{Clothing Deformation Model} ICP-style methods require good initialization to converge to the right local minimum. We observe that the underlying body pose can provide coarse information about body orientation and articulation as a good starting point. Thus we adopt a hierarchical deformation model for dynamic loose clothing: an outer layer of Linear Blend Skinning (LBS) $\mathcal W$ with respect to body pose, and an inner layer of deformation graphs $\mathcal E$ \cite{sumner2007embedded}: $\mathcal D(\boldsymbol \theta) = \mathcal W(\mathcal E(\mathbf M, \boldsymbol \theta), \boldsymbol\rho)$, where $\mathbf M$ is the template shape of the clothing defined in the canonical space of LBS. The underlying body pose $\boldsymbol\rho$ can be obtained from sparse RGB-D input by vision-based keypoint detection followed by inverse kinematics \cite{mehta2017vnect}, and it remains fixed during the N-ICP process. In this formulation, $\boldsymbol \theta$ is defined as the rotation and translation of deformation graph nodes, and we set its initialization to be the rest pose $\boldsymbol \theta^{(0)} = \mathbf 0$. Thanks to the global transformation and body articulation encoded in $\boldsymbol \rho$, $\mathcal D(\boldsymbol \theta^{(0)}) = \mathcal W(\mathbf M, \boldsymbol \rho)$ is close enough to the target $\mathbf P$ for N-ICP to converge nicely. This formulation also makes it efficient to perform per-frame tracking, preventing the failure caused by error accumulation in sequential tracking. The complete N-ICP process is illustrated in Fig.~\ref{fig:proposed-nicp}.

\section{Texel-Conditioned Clothed Avatars}
\label{sec:proposed-dense-avatar}

The N-ICP algorithm in Sec.~\ref{sec:proposed-nicp} can track large clothing dynamics, providing a good starting point for 
rendering the clothed body appearance. 
However, the underlying deformation model is designed to only capture large geometry deformations,
and does not model fine geometrical detail or appearance.
Therefore, in the next step, we develop a clothed avatar that can produce high-fidelity 
geometry and appearance when conditioned on both sparse RGB-D views as well as the output from the previous N-ICP stage. 
The critical question here is how to \textit{faithfully} reconstruct the appearance detail from
the sparse driving views for dynamic clothing.

We build upon the texel-conditioned avatars from Drivable Volumetric Avatars (DVA) \cite{remelli2022drivable}. DVA takes in driving signals of several RGB images mapped to texel-aligned 
features, as opposed to conditioned primarily on pose \cite{bagautdinov2021driving},
and predicts geometry and view-dependent appearance that can reproduce the full-body appearance. 
Formally, in DVA, the input feature $\mathbf F^b_I$ is the mean unwrapped image from multiple RGB driving 
views based on the skinned mesh $\mathbf W^b_{\boldsymbol\rho} = \mathcal W(\mathbf M^b, \boldsymbol \rho)$ 
of the body template $\mathbf M^b$. This unwrapping process can be written as 
\begin{gather}
    \mathbf F^b_I = \mathcal U(\{I_c\}, \mathbf W^b_{\boldsymbol\rho}),
    \label{eq:proposed-body-unwrap}
\end{gather}
where $\mathcal U$ denotes the unwrapping operation and $\{I_c\}$ denotes the set of driving images. The neural avatar $\mathcal A^b$ is a convolutional encoder-decoder that takes in $\mathbf F^b_I$, viewpoint $\mathbf v$ and body pose $\boldsymbol\rho$, and predicts the geometry corrective $\delta\mathbf G^b$ and appearance $\mathbf T^b$ with $[\delta \mathbf G^b, \mathbf T^b] = \mathcal A^b(\mathbf F^b_I, \mathbf v, \boldsymbol\rho)$.
The final geometry $\mathbf G^b$ is obtained by a function $\mathcal G$ that applies $\delta\mathbf G^b$ on top of the LBS mesh $\mathbf W^b_{\boldsymbol\rho}$ with a pre-defined coordinate transformation
\begin{gather}
\mathbf G^b = \mathcal G(\mathbf W^b_{\boldsymbol\rho}, \delta \mathbf G^b),
\label{eq:proposed-body-geo-update}
\end{gather}
which is then used to render\footnote{The geometry $\mathbf G$ and appearance $\mathbf T$ may take different forms depending on whether the Mixture of Volumetric Primitive (MVP) \cite{lombardi2021mixture} or mesh-texture formulation \cite{lombardi2018deep} is adopted as the rendering model, but the high-level concepts are similar and described here in a unified manner. The original DVA method~\cite{remelli2022drivable} uses MVP, while we use mesh and texture.} the output image together with the view-conditioned appearance $\mathbf T^b$.

\paragraph{Texel-Conditioned Clothing Avatars.} The DVA baseline, however, struggles to handle the 
large deformation of dynamic clothing. The root of the problem is that the LBS mesh $\mathbf W^b_{\boldsymbol\rho}$, 
which encodes only the skeleton-level motion, is too coarse to serve as the base geometry for dynamic clothing. 
The large deviation of the LBS mesh $\mathbf W^b_{\boldsymbol\rho}$ from the true clothing surface has two consequences: 
first, the unwrapping operation in Eq.~(\ref{eq:proposed-body-unwrap}) cannot effectively capture the appearance detail; 
second, it places a heavy burden for the network $\mathcal A^b$ to predict a large offset $\delta\mathbf G^b$ to 
update the geometry in Eq.~(\ref{eq:proposed-body-geo-update}).

One of the key ideas of our clothed avatar model is to use the non-rigid tracking result $\mathbf D = \mathcal D(\boldsymbol\theta^{(N)})$ from the final N-ICP iteration as the starting point. Because $\mathbf D$ is already well-aligned with the clothing surface, we can obtain better texel-aligned features with more appearance detail from the unwrapping operation $\mathbf F_I = \mathcal U(\{I_c\}, \mathbf D)$.
In order to further guide the estimation of the geometry corrective using the driving depth images, we also provide a 
``depth offset'' feature $\mathbf F_D$ as input. 
For each pixel $[x, y]$ in camera $c$ with depth value $D_c$, we associate it with the rendered depth $d_c$ from the tracked geometry $\mathbf D$ at the same pixel location and compute the offset as
\begin{gather}
    O_c[x, y] = \mathbf R_c (d_c[x, y] - D_c[x, y]) \mathbf K_c^{-1} \begin{bmatrix} x & y & 1 \end{bmatrix}^T + \mathbf t_c,
\end{gather}
where $\mathbf K_c$ is the camera projection matrix, and $\{\mathbf R_c, \mathbf t_c\}$ are the rigid transformation from each camera frame to a unified body root coordinate frame. The depth offset feature is then a 3-channel average tensor obtained by unwrapping $\{O_c\}$ for each driving camera $c$, $\mathbf F_D = \mathcal U(\{O_c\}, \mathbf D)$.
Thus, $\mathbf F_D$ can be regarded as the offset to be corrected on top of $\mathbf D$ in order to match the sensor depth for each UV location. Our clothing model takes in the texel-aligned features $\mathbf F_I$ and $\mathbf F_D$ as well as the viewpoint $\mathbf v$, and predicts the geometry corrective $\delta \mathbf G$ and texture $\mathbf T$ with $[\delta \mathbf G, \mathbf T] = \mathcal A(\mathbf F_I, \mathbf F_D, \mathbf v)$.
Finally, the geometry $\mathbf G$ and texture $\mathbf T$ are used to render the output image. Following \cite{xiang2021modeling}, the clothing is modelled as a separate layer from the underlying body avatar. The body avatar can be conditioned on body pose $\boldsymbol\rho$ or additionally on texel-aligned features \cite{remelli2022drivable}.
To train the model, besides the standard image reconstruction loss and mesh regularization, we use a part segmentation loss and an ID-MRF perceptual loss \cite{wang2018image}, which are detailed in the supplementary document.

%% file: 4-results.tex
\section{Results}


\subsection{Capture Setup and Detail}
\label{sec:proposed-results-setup}

We capture a total of three sets of garments: (1) a red dress with a short full skirt; (2) a flared, short skirt\footnote{The paired upper-body garments are tight and modelled in the body layer.} in floral pattern with a bottom ruffle; (3) a loose T-shirt and a long skirt. The training data are captured in a multi-view capture studio equipped with roughly 150 RGB cameras. This dense capture setup enables us to use Multi-View Stereo (MVS) to reconstruct geometry, which can be rasterized to depth and used as input to drive our avatars. We split out a small segment of each sequence for testing purpose and use the multi-view captured images as ground truth for evaluation.

To demonstrate the application of our method for telepresence, we additionally capture the subjects with the same garments in a novel environment with different background and illumination. It is designed to be a simpler capture setup, where nine Kinect RGB-D cameras are deployed, and we select three cameras as driving input to our model. We also split the data in the novel environment for fine-tuning and testing.
%
%

\subsection{Evaluation of Neural Iterative Closest Point}
\label{sec:proposed-results-nicp}

We conduct an evaluation on the N-ICP algorithm using the dress sequence. We use the ground truth from offline registration by non-rigid ICP from \cite{xiang2022dressing}. For this evaluation, we adopt the same input as our full method: a point cloud fusing the depth maps from three driving views. We report the evaluation metric of the mean squared point-to-triangle distance for both directions from prediction to ground-truth and from ground truth to prediction.

\begin{figure}[t]
    \centering
    \includegraphics[width=\linewidth]{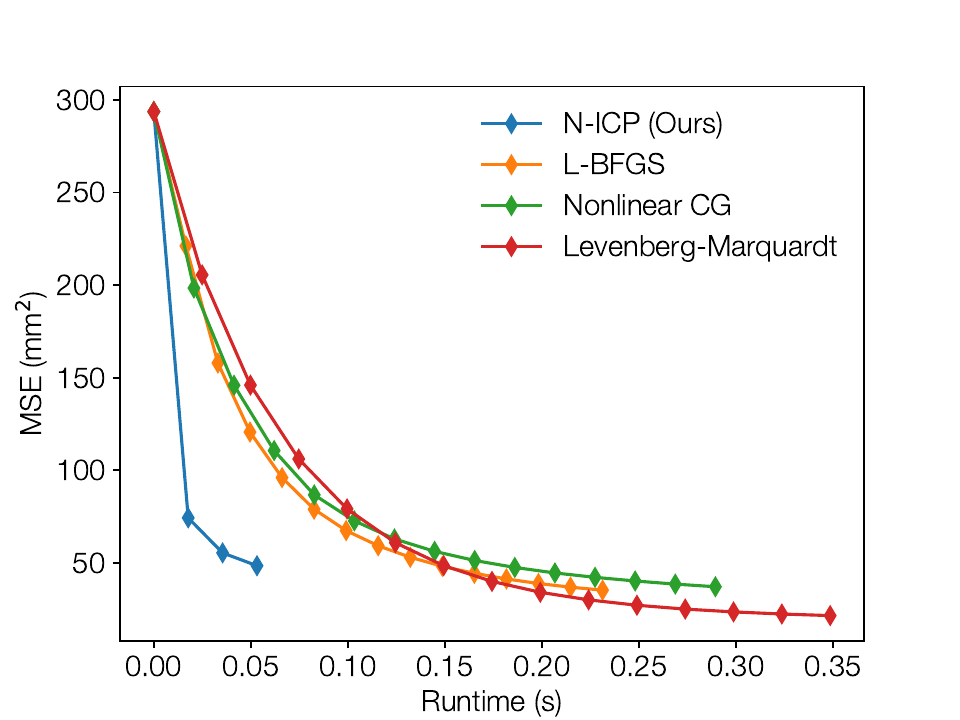}
    \caption{Comparison between our method and classical non-rigid ICP with different types of optimization solvers. For each method, we plot the runtime in seconds vs. the Mean Squared Surface Error (MSE) in $\si{\milli\metre\squared}$ for the surface tracking results. The square markers on the curves denote individual steps in the optimization.}
    \label{fig:proposed-eval-nicp-steps}
\end{figure}

We compare N-ICP with classical optimization solvers, including L-BFGS and nonlinear Conjugate Gradient with strong Wolfe linear search implemented in PyTorch-Minimize \cite{Feinman2021}, and Levenberg-Marquart implemented in Theseus \cite{pineda2022theseus}. All the methods are implemented in CUDA PyTorch with analytically computed derivatives. The results are shown in Fig.~\ref{fig:proposed-eval-nicp-steps}. Our method converges at a faster rate than the baseline methods, including both the gradient-based methods (L-BFGS, CG and ours) and the more complicated Jacobian-based Levenberg-Marquardt method. For ablation studies on the formulation of our N-ICP network, please refer to our supplementary document.


\subsection{Full Method Evaluation}
\label{sec:proposed-results-studio}

\begin{table}[t]
\centering
\caption{Quantitative comparison with various baselines and ablation studies using the loose T-shirt and long skirt sequence captured in the multi-view studio. The metrics are computed on the whole rendered images with a plain background from two different views. We compare with DVA \cite{remelli2022drivable} and its two-layer variant, ENeRF \cite{lin2022efficient}, KeypointNeRF \cite{mihajlovic2022keypointnerf}, and sensing-based baselines. Ablation studies are also included.}
\begin{tabular}{ L{3.1cm}  C{0.9cm}  C{0.9cm}  C{0.9cm}  C{1.0cm} }
    \toprule
    Methods & $L_1\downarrow$ & PSNR$\uparrow$ & SSIM$\uparrow$ & LPIPS$\downarrow$ \\
    \midrule
    DVA & 2.89 & 27.96 & 0.9218 & 0.0932 \\
    DVA (Two-layer) & 2.90 & 28.03 & 0.9219 & 0.0912 \\
    \midrule 
    ENeRF (Masked) & 3.07 & 27.21 & 0.9130 & 0.0772 \\
    KeypointNeRF & 3.14 & 27.42 & 0.9111 & 0.1123 \\
    \midrule
    Depth-based warping & 3.22 & 26.94 & 0.9171 & 0.1111 \\
    + LookinGood UNet & 2.27 & 29.91 & 0.9382 & 0.0865 \\
    \midrule 
    Ours w/o N-ICP & 2.29 & 29.04 & 0.9299 & 0.0755 \\
    Ours w/o depth offset & 2.19 & 29.35 & 0.9327 & 0.0728 \\
    Ours w/o part loss & 2.11 & 29.63 & 0.9354 & 0.0724 \\
    \midrule
    Ours full & \textbf{1.93} & \textbf{30.25} & \textbf{0.9404} & \textbf{0.0686}  \\
    \bottomrule
\end{tabular}
\label{table:proposed-eval-socio}
\end{table}

In this section, we present evaluation and comparison using high-quality data from the dense multi-view capture studio. For comparison with most methods, we report quantitative results on the the challenging loose T-shirt and long skirt sequence in in Tab. \ref{table:proposed-eval-socio}, except for pose-driven clothed avatars because the tucked T-shirt is extremely challenging to simulate in \cite{xiang2022dressing}. Therefore for this category we report results on the skirt with floral pattern shown in Tab. \ref{table:eval-cca-da}.



\paragraph{Comparison w/ DVA \cite{remelli2022drivable}.} Although DVA adopts the Mixture of Volumetric Primitive (MVP) \cite{lombardi2021mixture} formulation that can render any structure in theory, the spatial arrangement of the primitives relies on the guidance of base body mesh as initialization. Both the original method and its variant with body and clothing layers are guided by LBS transformation, which is too coarse to capture the motion of loose clothing.
Qualitative comparison is shown on the left of Fig.~\ref{fig:combine-dva-nerf}.

\paragraph{Comparison w/ NeRF-based methods.} NeRF has been used to model dynamically moving human appearance, but many of them requires per-scene/frame fitting \cite{isik2023humanrf,shao2023tensor4d} and are not directly applicable to our telepresence setting. We only compare our method with representative approaches of generalizable NeRF that directly take in sparse image input. ENeRF \cite{lin2022efficient} focuses on the setting of interpolating a target view from several \emph{most nearby} source views without human prior, but does not perform well in our setting due to the difficulty in associating views given wide camera baselines. KeypointNeRF \cite{mihajlovic2022keypointnerf} provides relative encoding of body keypoints as an additional input feature for radiance prediction but cannot handle clothing motion that does not closely follow the body motion. Qualitative comparison is shown on the right side of Fig. \ref{fig:combine-dva-nerf}.

\begin{figure}[t]
    \centering
    \includegraphics[width=\linewidth]{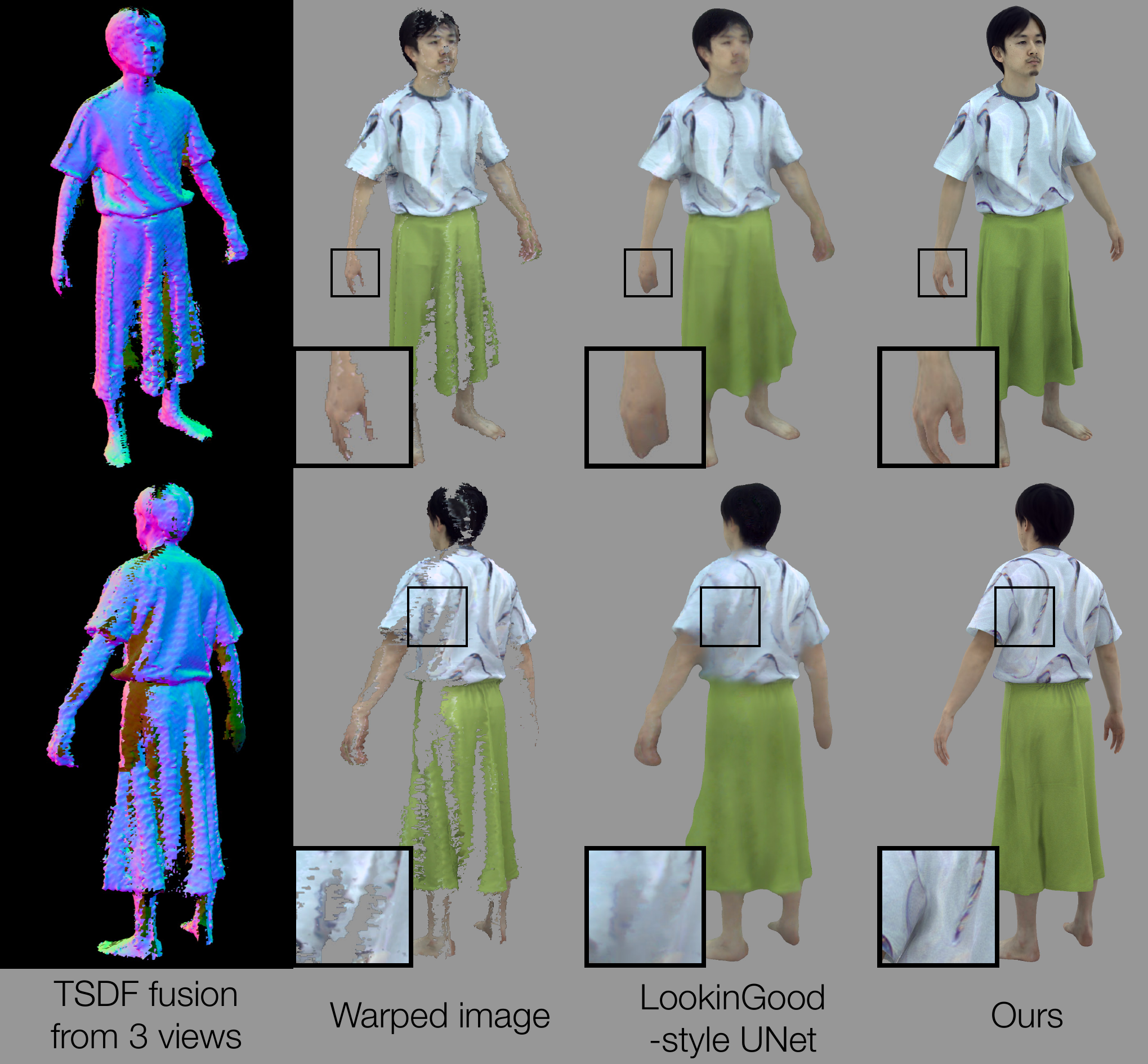}
    \caption{Comparison with basic sensing-based baselines. Given the depth input from 3 views, we run TSDF fusion \cite{curless1996volumetric,dong2022ash} to obtain a proxy geometry (first column), and then warp the driving RGB images to the target view (second column). We train a UNet following LookinGood \cite{martin2018lookingood} to inpaint the missing regions (third column).}
    \label{fig:comparison-sensing}
\end{figure}

\paragraph{Comparison w/ sensing-based baselines.} Most sensing-based approaches \cite{dou2017motion2fusion,yu2021function4d,martin2018lookingood,shao2022floren} are proprietary and tightly integrated with their capture system without available open-source implementations. Therefore, we rather compare our method with baselines that can be implemented with moderate effort, listed in the third group of results in Tab.~\ref{table:proposed-eval-socio}, including image warping based on TSDF fusion and further applying a U-Net follwing the idea of LookinGood \cite{martin2018lookingood}. For detail please refer to our supplementary document.

\begin{figure}[t]
    \centering
    \includegraphics[width=\linewidth]{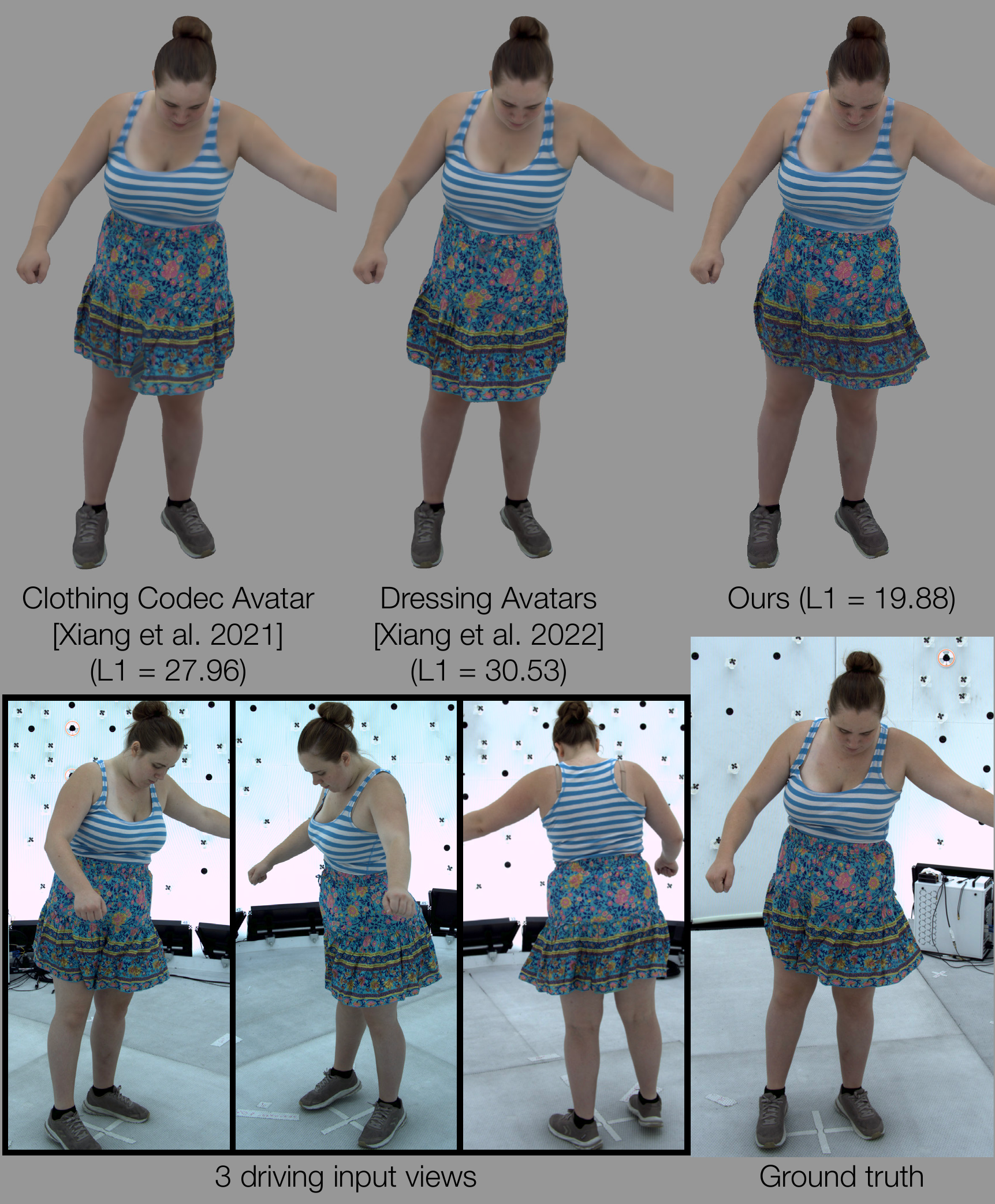}
    \caption{Qualitative comparison with Clothing Codec Avatars \cite{xiang2021modeling} and Dressing Avatars \cite{xiang2022dressing}. The results are shown on the top row; the input views and ground truth are shown on the bottom. The $L_1$ error in the skirt region is reported beside the names of the methods.}
    \label{fig:comparison-prev-cloth-work}
\end{figure}

\begin{table}[t]
\centering
\caption{Quantitative comparison with Clothing Codec Avatars \cite{xiang2021modeling} and Dressing Avatars \cite{xiang2022dressing} on the skirt sequence. The metrics are computed in the skirt region.}
\begin{tabular}{ L{2.5cm}  C{0.9cm}  C{0.9cm}  C{0.9cm}  C{1.0cm} }
    \toprule
    Methods & $L_1$$\downarrow$ & PSNR$\uparrow$ & SSIM$\uparrow$ & LPIPS$\downarrow$ \\
    \midrule
    Clothing CAs & 25.46 & 17.81 & 0.2431 & 0.507 \\
    Dressing Avatars & 27.95 & 17.01 & 0.1818 & 0.486\\
    Ours & \textbf{15.67} & \textbf{21.99} & \textbf{0.6527} & \textbf{0.203} \\
    \bottomrule
\end{tabular}
\label{table:eval-cca-da}
\end{table}

\paragraph{Ablation studies.} We show ablation studies on several components of our framework and the results are shown
in Fig.~\ref{fig:comparison-ablation} and the fourth group of Tab.~\ref{table:proposed-eval-socio}. 
First, the initial tracking by N-ICP provides a basis for the whole framework to estimate the correct clothing geometry and appearance. Without this component, the initialization by only body pose is too coarse and leads to obvious artifacts. Second, we demonstrate that the additional depth offset input to the encoder-decoder allows our method to predict more accurate overall clothing shape. Third, we verify that the part segmentation loss helps to produce correct body-clothing boundary. Last, we compare the results with and without ID-MRF loss, which preserves the high-frequency texture detail in our output when the predicted geometry is not completely photometrically aligned with the driving images.

\paragraph{Comparison w/ pose-driven avatars.} A major motivation of our framework is that the underlying body motion does not contain enough information to fully determine the loose clothing states. We validate this intuition by comparison with Clothing Codec Avatars \cite{xiang2021modeling} and Dressing Avatars \cite{xiang2022dressing} shown in Fig.~\ref{fig:comparison-prev-cloth-work} and Tab.~\ref{table:eval-cca-da}. Clothing Codec Avatars struggle to learn the mapping from a sequence of body pose to large clothing dynamics. Dressing Avatars produce clothing motion that looks \emph{realistic} with the help of physics-based simulation, but its output is not \emph{faithful} to the actual motion because of the lack of an efficient approach for estimating underlying physical parameters for simulation. Our method utilizes the driving signal from sparse RGB-D input to achieve faithful clothing telepresence, which is verified by the low error on the evaluation metrics in Tab. \ref{table:eval-cca-da}. Please refer to the supplementary document for a more detailed discussion on the difference in their formulation (e.g. supervised vs. self-supervised).

\subsection{Results in the Novel Capture Environment}
\label{sec:proposed-results-stageroom}

For the novel capture environment, we test our model in two different scenarios: 
without and with fine-tuning, both shown in Fig.~\ref{fig:stage}.
The first scenario refers to a direct application of the model trained from the dense capture studio
to the same subject but in the novel environment. 
For this scenario, our model needs to handle the difference in the input RGB-D images between the training
and testing environments, such as illumination and sensor setup. 
Note that our model is trained to be robust to these variations in the input and preserve
the appearance style from the original training data in our output (see supplementary document), but with the unseen body and clothing motion captured in the novel environment. 
This experiment demonstrates the ability of our method to directly generalize to a novel environment.

In the second scenario, we test our model after fine-tuning it with the training data captured in the new environment. 
Note that we use the same three Kinects for both driving input and ground truth supervision. 
During fine-tuning, the output of our model adapts to the new appearance caused by the illumination and sensor
specification in the environment, as well as the change in body over time such as hair style. We test our model with a even more aggressive change in the body-layer appearance from the tank-top to the green capture suit in the second row of Fig.~\ref{fig:stage}.
The fine-tuning step can be viewed as an option to further boost the model output quality if 
time and computation budgets permit.

%% file: 5-conclusion.tex
\section{Conclusion}

We presented a framework for building photorealistic full-body avatars that can be driven by sparse RGB-D 
inputs and faithfully reproduce the motion of loose clothing. 
Our method accurately reconstructs challenging clothing appearance of the subjects, thus tackling a major drawback of 
existing pose-driven avatars.
%
As per limitations, our model is still person- and garment-specific and cannot handle clothing motion 
that falls far outside of the deformation space of the deformation graph model, such as topology change.
Interesting future directions would be to extend our method to multi-identity setting, and develop a formulation that 
can handle more generic garment categories, e.g. with implicit representations.


\bibliographystyle{ACM-Reference-Format}
\bibliography{local}

\clearpage

\begin{figure*}[t]
    \centering
    \includegraphics[width=0.925\linewidth]{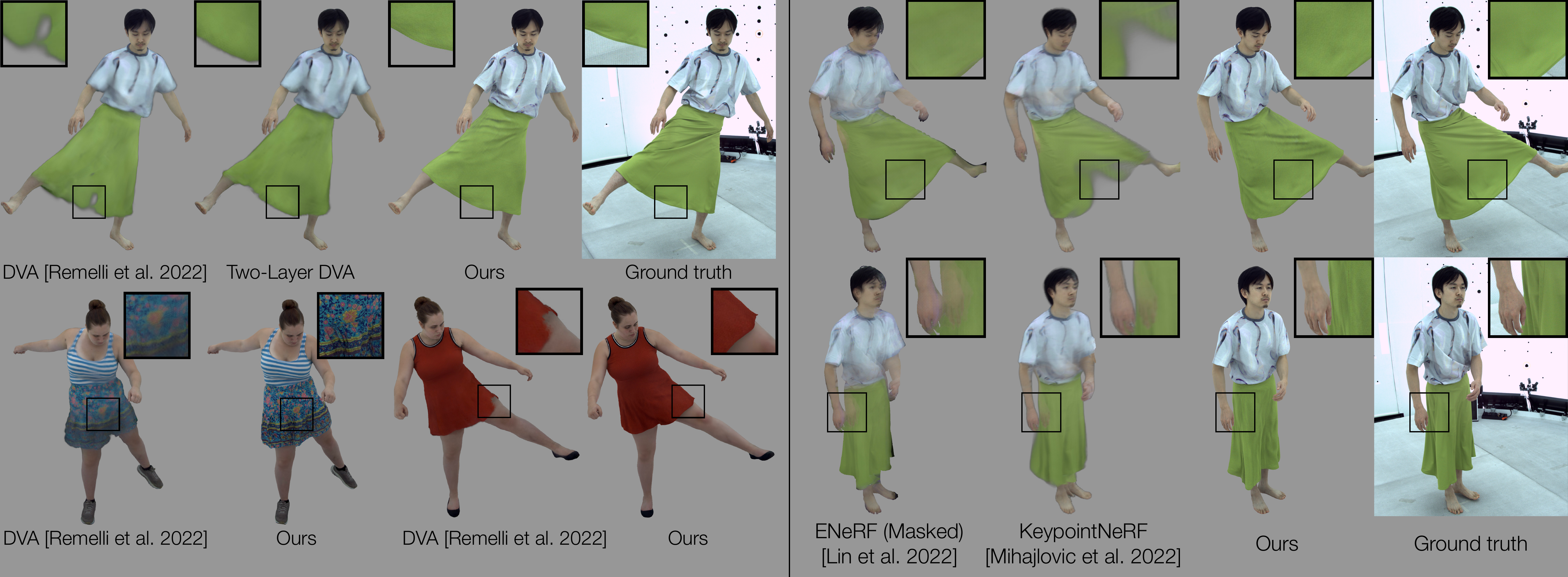}
    \caption{\textbf{Left}: comparison with Drivable Volumetric Avatars (DVA) \cite{remelli2022drivable} and its two-layer variant. \textbf{Right}: comparison with NeRF-based methods \cite{lin2022efficient,mihajlovic2022keypointnerf}. We mask out the black background of results in ENeRF \cite{lin2022efficient} with ground truth segmentation.}
    \label{fig:combine-dva-nerf}
\end{figure*}

\begin{figure*}[t]
    \centering
    \includegraphics[width=0.925\linewidth]{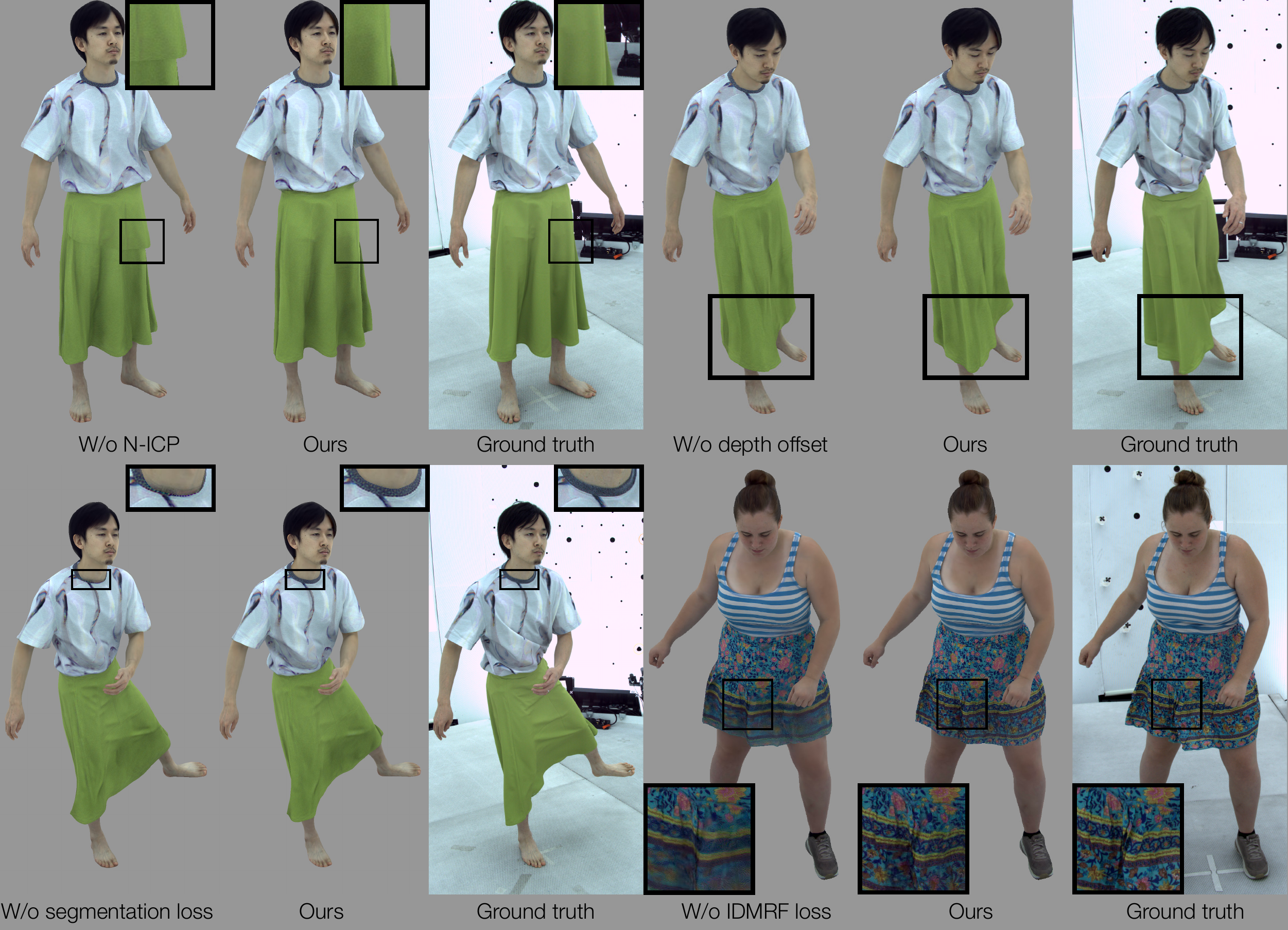}
    \caption{Ablation studies on components in our framework, including N-ICP, depth offset as input to the encoder-decoder, part segmentation loss, and ID-MRF loss. Each result is shown in comparison to our full output and the ground truth.}
    \label{fig:comparison-ablation}
\end{figure*}

\begin{figure*}[t]
    \centering
    \includegraphics[width=\linewidth]{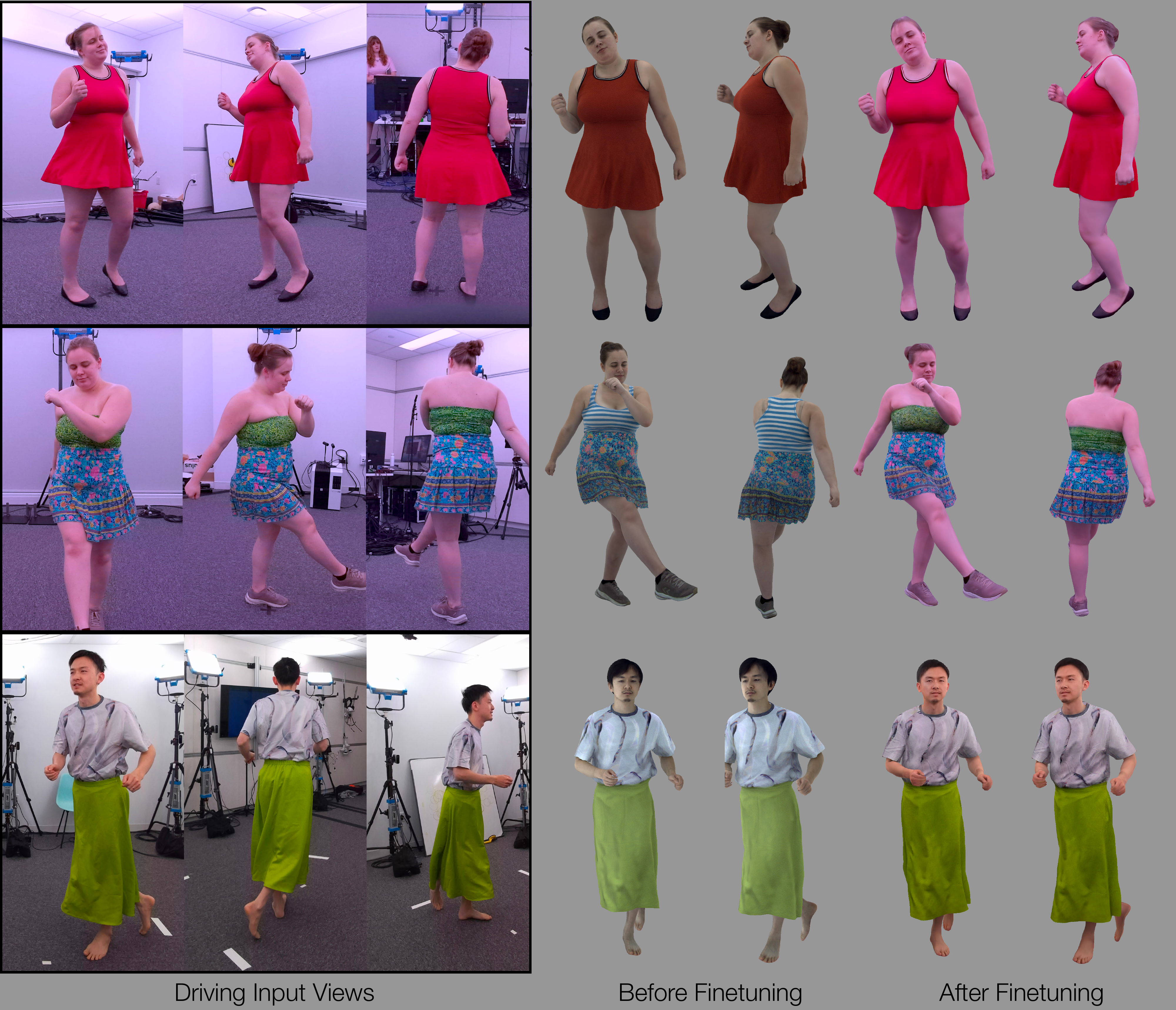}
    \caption{Testing results in the novel capture environment. On the left we show in the input RGB-images. We show our output from 2 different viewpoints both before and after fine-tuning. These frames are not seen during fine-tuning. The subject on the second row wears different upper-body garments between the captures in the original and new environment. The tank top is preserved in the body layer before fine-tuning and adapted after that.}
    \label{fig:stage}
\end{figure*}